\documentclass[a4paper,11pt]{article}
\pdfoutput=1 

\usepackage{jheppub} 

\usepackage[T1]{fontenc} 

\title{\boldmath Obstruction of black hole singularity by quantum field theory effects}

\author[a,b]{Jahed Abedi}
\author[a,b]{and Hessamaddin Arfaei}
\affiliation[a]{Department of Physics, Sharif University of Technology,\\
P.O. Box 11155-9161, Tehran, Irany}
\affiliation[b]{School of Particles and Accelerators,\\ Institute for Research in Fundamental Sciences (IPM), \\
 P.O. Box 19395-5531, Tehran, Iran}

\emailAdd{jahed\_abedi@physics.sharif.ir}
\emailAdd{arfaei@ipm.ir}

\abstract{We consider the back reaction of the energy due to quantum fluctuation of the background fields considering the trace anomaly for Schwarzschild black hole. It is shown that it will result in modification of the horizon and also formation of an inner horizon. We show that the process of collapse of a thin shell stops before formation of the singularity at a radius slightly smaller than the inner horizon at the order of $(c_{A}\frac{M}{M_p})^{1/3}l_p$. After the collapse stops the reverse process takes place. Thus we demonstrate that without turning on quantum gravity and just through the effects the coupling of field to gravity as trace anomaly of quantum fluctuations the formation of the singularity through collapse is obstructed. An important consequence of our work is existence of an extremal solution with zero temperature and a mass which is lower bound for the Schwazschild solution. This solution is also the asymptotic final stable state after Hawking radiation.}

\begin{document} 
\maketitle
\flushbottom

\section{\label{Introduction}Introduction}

The question of formation of Black hole singularity as a result of collapse of shell or ball of matter, despite many years of study is still unresolved. In a classical scenario a spherical shell or a ball of matter collapses and passing the Schwarzschild radius moves toward the singularity as far as classical Einstein equation is valid i.e. where quantum effects take the lead. It bounds the fate of singularity to a theory for quantum gravity. Of course one need not go this far to revise the scenario since at least Hawking radiation will certainly change the effective mass and will  affect the final outcome but does not remove the singularity at the center. The question is whether there are other effects that  can result in big change in the picture to the extent of obstructing the formation of the singularity. It is shown in \cite{Paranjape:2009ib,Frolov:2014wja,Bardeen:2014uaa} that Hawking radiation is not strong enough to prevent formation of neither the horizon nor the singularity but will delay the process. Also in a previous work \cite{Abedi:2013xua} the authors have shown that in the context of dilatonic theory the primordial black holes are strongly modified due to Hawking radiation but the horizon is formed although adiabatically changes. It is shown that  other singularities such as big bang singularity \cite{Ashtekar:2006rx} are  modified considerably when quantum gravitational effects are considered.
 Similar considerations  were applied to Black holes in several articles arguing that the collapse shall stop before the singularity is formed \cite{Bambi:2013gva,Bambi:2013caa,Rovelli:2014cta,Bardeen:2014uaa,Mersini-Houghton:2014zka,Mersini-Houghton:2014cta,Modesto:2004xx,Torres:2014gta,Torres:2015aga,Chakraborty:2015nwa,Taves:2014laa,Vaz:2014era,Culetu:2014tqa,Hayward:2005gi,Kawai:2013mda,Kawai:2014afa,Barcelo:2014npa,Barcelo:2014cla}. They either  rely on arguments based on quantum gravity or Hawking radiation. In particular in the context of loop quantum gravity Rovelli et al \cite{Rovelli:2014cta} have argued that quantum corrections can impede the collapse of a matter to the extent that it be stopped much before the black hole radius reaches the Planck length by a factor $\left(\frac{M}{M_{p}}\right)^{\frac{1}{3}}$, hence preventing the formation of the singularity, forming what is called Planck Star. In addition, in the context of weakly nonlocal quantum gravity the similar behaviour is expected \cite{Frolov:2015bta,Frolov:2015bia,Modesto:2014lga,Modesto:2011kw}. The analysis follows the previous observation by Ashtekar and coworkers \cite{Ashtekar:2006rx} who suggested formation of a bounce radius based on gravitational quantum effects for big bang when we go back in time. There has also been attempts to find a transition from black holes to white holes as the effect of quantum gravity as the collapse process \cite{ Barcelo:2014npa, Barcelo:2014cla}. 

There are also other arguments in support of the resolution of the singularity. Theoretical arguments suggesting that the information is carried out by black hole radiation leads to phenomenon such as firewall \cite{ Almheiri:2012rt, Braunstein:2009my}. But if information is trapped inside, the final stage of the black hole evaporation carries the amount of information that would hardly be compatible with the Planck length \cite{ Giddings:1992ff, Page:1993wv, Hossenfelder:2009xq}. In order to resolve this problem Giddings suggested that the size of the black hole at the final stage of its evaporation must be much larger than the Planck length \cite{ Giddings:1992hh}. Based on these arguments and as a consequence of loop quantum cosmology Rovelli and Vidotto \cite{ Rovelli:2014cta} suggested that instead of singularity, a new object will form, Planck Star at a so called bounce radius much larger than the Planck length which may help to provide a resolution of the BH information Paradox.

The question we consider in this note is whether Quantum Field Theory (QFT) effects can substantially modify the process of collapse. The particular effect we consider is the change of the vacuum energy of fields living in the black hole background. Since the fields are coupled to the background geometry, we explore whether the zero point energy of the field(s) can lead to significant correction to the Einstein equation and modify the solution (in our case the Schwarzschild solution)  and the process of collapse in particular the formation of the singularity. 
Here, as the matter proceeds during the collapse, zero point or vacuum energy of the fields become more and more relevant to the fate of the collapsing shell since it depends on curvature through higher power terms which modify the Einstein equation giving rise to an inner horizon, inside the Schwarzschild radius, much larger than the Planck length. The zero point energy resembles the Casimir energy inserting an (outward) pressure on the boundary of the collapsing shell opposing the process of collapse. Is this pressure strong enough to stop it?

We can put the question in a different way; is the energy transferred to the environment large enough to modify the Schwarzschild solution significantly and stop the collapse? We answer this question affirmatively by taking the contribution of the zero point energy due to trace anomaly. We will take the advantage of the trace anomaly which is exact going through the regions beyond the validity of semi-classical approximations at the center of the black hole. In fact we shall show that the collapsing shell first enters the inner horizon, much larger than the Planck length, then stops at a so called bounce radius $R_{B}$, before the formation of the singularity, where all of the relativistic kinetic energy of the shell is transferred to the background vacuum. The shell reaches and stops at the turning point, bounce radius, which is also much larger than the Planck length and after that the process is reversed.

We also consider the geodesics of an infalling particle in the quantum corrected metric and show that the geometry acts like an effective potential preventing it from reaching the center. Then the particle returns inside the inner horizon and moves in outward direction.

The field theory we consider is a massless free theory (non-interacting) coupled only to gravity. By considering the trace anomaly we are able to extract trace part of the energy momentum tensor, the pressure due to vacuum fluctuation of the fields.

We are able to show that QFT effects through trace anomaly can result in resolution of the BH singularity for the process of collapse and QG need not be turned on. Interestingly, our considerations give the same behaviour for the bounce radius found for the Planck star in the context of quantum gravity \cite{Rovelli:2014cta}. Although we only consider the effect of trace anomaly that is a part of the quantum correction, we expect that other quantum effects strengthen the phenomena. This argument is also supported by Israel and Poisson \cite{0264-9381-5-12-002} where they state that the vacuum polarization can regularize the metric replacing the
 center by a sub-Planckian core extracting the possible form of $g^{rr}$.

A suitable model is provided by the collapse of an infinitely thin spherical shell from a homogeneous ideal fluid. This model is adopted for simplicity and provides the essence of the phenomena.

We follow the standard treatment of Israel \cite{Israel:1966} to match the internal and external solutions,  using the geometrical notion of extrinsic curvature of the hypersurface embedded in the surrounding space-time (see also \cite{Kuchař:1968, Giddings:2001ii, poisson}).

The vacuum expectation value of the energy-momentum tensor can be divided into anomalous part and traceless part. The anomalous part is calculated exactly by many authors \cite{PhysRevD.33.2840, Capper, Deser197645, Duff1977334, Duff:1993wm, PhysRevD.39.2993} (see also \cite{Davies, Mottola:2010gp}), while the other part is given by approximate methods \cite{Belokogne:2014ysa,PhysRevD.29.1057}.

It is worth mentioning that there are pioneer attempts to account for effects of trace (conformal) anomaly in solution of Einstein equation by Bardeen \cite{ Bardeen:2014uaa}, Hawking effect \cite{ PhysRevD.15.2088} by Christensen and Fulling and \cite{Kawai:2014afa} by Kawai, and Yokokura, and \cite{Shapiro:2008sf,Balbinot:1999vg,Balbinot:1999ri} by Balbinot, and, Fabbri, and Shapiro as well as effects in the stellar context \cite{Elizalde:1999dw, Bytsenko:1998md, Nojiri:1998ph, Nojiri:1999pm} by Odintsov, Nojiri, Elizalde, and Bytsenko and holographic black holes \cite{Casadio:2004nz, Casadio:2005if} by Casadio and Germani.

In this setup energy density of the required fields $\rho=-\langle T_{t}^{t} \rangle$ plays significant role (fields with positive energy density obstruct the formation of the singularity \cite{0264-9381-5-12-002}). Based on \cite{Chakraborty:2015nwa,Singh:2014paa} this energy density for freely falling observer is positive and starts to rise and ultimately diverges as the singularity is approached. In the cosmological backgrounds similar effect takes place \cite{PhysRevD.11.3370,PhysRevD.14.3304}. It worth to add that the positive vacuum energy density is among the assumptions employed in the construction of  the non-singular black holes \cite{Hayward:2005gi, 0264-9381-5-12-002}. In this paper all the particles obstruct the formation of the singularity\footnote{Although for vector particles $c_{A}$ is negative, higher order corrections stay positive inside the core such that the final result does not change.}. On the grounds that for particles dealt with in this paper the trace anomaly cannot be cancelled \cite{PhysRevD.39.2993}, its contribution to the collapse cannot be disregarded. Since massive particles have less important contributions in zero point energy, we ignore their effects and limit ourself to massless cases.

Several scenarios can be envisaged as the result of Hawking radiation. They all point to merging of the inner and outer horizons \cite{Hayward:2005gi}. The detail and the timing of the effect is central to the scenario. In this paper we have not addressed the question.

The setup we consider is the same setup that leads to formation of Schwarzschild black hole i.e. no conserved charge is taken to exist and the background is taken to be empty. This is in contrast to realistic collapse that definitely carries baryon charge and live in non-empty surrounding. The dust and background radiation \cite{Firouzjaee:2014zfa,Firouzjaee:2011hi} is not considered in the setup and  issues will be addressed elsewhere.  Note that such realistic conditions will help the impeding of the collapse and will strengthen the final result. We make the standard assumptions to analyse the back reaction of the change in the energy momentum tensor induced by the coupling of the fluctuations of the Quantum fields to gravity \cite{Paranjape:2009ib,Brout:1995rd}; the metric inside the shell is Minkowski metric, the collapse is spherically symmetric with no rotation, and the Einstein equation is modified to the form,
\begin{eqnarray}
G_{\mu \nu}=8 \pi ( t_{\mu \nu}+\langle T_{\mu \nu} \rangle). \label{eq1.1}
\end{eqnarray}
In the next section we first obtain the quantum corrected metric arising from the trace anomaly and existence of extremal solution and a lower bound for mass of the black hole. Then we derive the trajectory of an infalling particle in this background. Lastly we present the main results by determining the gravitational collapse of a shell in this background. We also obtain trajectory of the shell, and bounce radius and discuss about the stability of the final state of the black hole.
Finally, in the last section we end with concluding remarks and future plans.


\section{\label{Gravitational collapse considering vacuum energy effects}Gravitational collapse considering vacuum energy effects}

We assume and shall analyse the  Einstein equation (\ref{eq1.1}) generalized to include the zero point energy resulting from the fluctuation of the fields with the flat space contribution subtracted,  $ \langle T_{\mu \nu} \rangle $.
As in the standard case we concentrate on the  formation of Schwarzschild black hole from the collapse of a  rotationally  symmetric spherical thin shell of radius $R$. According to the classical picture \cite{Israel:1966,Kuchař:1968, Giddings:2001ii, poisson} inside the shell we will have a flat and outside a Schwarzschild metric ($r>R$).

Note that at the classical level this background is  Ricci flat ($\mathcal{R}_{\mu \nu} = \mathcal{R}=0$).
\begin{eqnarray}
ds^{2}=-(1-\frac{2M}{r}) dt^{2} + \frac{dr^{2}}{(1-\frac{2M}{r})} + r^{2}(d\theta^{2}+ \sin^{2}\theta d\phi^{2})
\end{eqnarray}
where M is ADM mass of the black hole.

We will study the backreaction of the QFT energy released to the part of the space whose metric changes from flat to Schwarzschild to the collapse process.
To go beyond the classical approximation and include the QFT effects we take  $\langle T_{\mu \nu} \rangle$ in the background of the classical solution, and solve for the correction to the new equation. Note that the added term in the equation changes as the collapse goes on.
 
We  calculate $\langle T_{\mu \nu} \rangle$ from the trace anomaly of the fields coupled to the gravity.
In a general four dimensional curved space-time, the trace anomaly takes following form \cite{Mottola:2010gp},
\begin{eqnarray}
\langle T_{\rho}^{\rho} \rangle = \frac{\hbar}{32 \pi} \{ (c_{A}+c^{\prime}_{A}) ( \mathcal{F} + \frac{2}{3} \square \mathcal{R} ) - c^{\prime}_{A} \mathcal{E} + c^{\prime \prime}_{A} \square \mathcal{R} \}, \label{eq2.2}
\end{eqnarray}

where
\begin{eqnarray}
\mathcal{E}=~^\ast\!\mathcal{R}_{\mu \nu \rho \sigma} ~^\ast\!\mathcal{R}^{\mu \nu \rho \sigma}  \nonumber \ \ \ \ \ \ \ \ \ \ \ \ \ \ \ \ \ \ \ \  \\ 
\ \ = \mathcal{R}_{\mu \nu \rho \sigma} \mathcal{R}^{\mu \nu \rho \sigma}- 4 \mathcal{R}_{\mu \nu} \mathcal{R}^{\mu \nu} + \mathcal{R}^{2}
\end{eqnarray}
and
\begin{eqnarray}
\mathcal{F}=\mathcal{C}_{\mu \nu \rho \sigma} \mathcal{C}^{\mu \nu \rho \sigma}  \nonumber \ \ \ \ \ \ \ \ \ \ \ \ \ \ \ \ \ \ \ \ \ \ \ \ \ \  \\ 
\ \ = \mathcal{R}_{\mu \nu \rho \sigma} \mathcal{R}^{\mu \nu \rho \sigma}- 2 \mathcal{R}_{\mu \nu} \mathcal{R}^{\mu \nu} +\frac{ \mathcal{R}^{2}}{3}
\end{eqnarray}
with $\mathcal{R}_{\mu \nu \rho \sigma}$ the Riemann curvature tensor, $~^\ast\!\mathcal{R}_{\mu \nu \rho \sigma} = \epsilon_{\mu \nu \alpha \beta}\mathcal{R}^{\alpha \beta}{}_{\rho \sigma} /2$  its dual, and $\mathcal{C}_{\mu \nu \rho \sigma}$ the Weyl conformal tensor, where 
\begin{eqnarray}
c_{A}=\frac{1}{90\pi} (n_{0} + \frac{7}{4} n^{M}_{1/2} + \frac{7}{2} n^{D}_{1/2} - 13 n_{1} + 212n_{2}) \label{eq2.5}
\end{eqnarray}
and
\begin{eqnarray}
c^{\prime}_{A}=\frac{1}{90\pi} (\frac{1}{2} n_{0} + \frac{11}{4} n^{M}_{1/2} + \frac{11}{2} n^{D}_{1/2} + 31 n_{1} + 243n_{2}) \label{eq2.6}
\end{eqnarray}
\cite{PhysRevD.39.2993, Parker1979, Christensen1978571} are independent of the state in which the expectation value of the stress tensor is computed, and do not depend on any ultraviolet short distance cutoff \cite{PhysRevD.33.2840, Capper, Deser197645, Duff1977334, Duff:1993wm}. Here $n_{s}$ represents the number of particles for a particular spin $s$. For fermions superscripts M and D indicate the real (Majorana) spinors and complex (Dirac) spinors respectively.

The covariant form of the energy momentum tensor in Einstein equation (\ref{eq1.1}) is given by,
\begin{eqnarray}
\langle T_{\mu \nu} \rangle 
 =  \frac{\hbar}{64 \pi} \{ (c_{A}+c^{\prime}_{A}) ( \mathcal{F} + \frac{2}{3} \square \mathcal{R} ) - c^{\prime}_{A} \mathcal{E}+c^{\prime \prime}_{A} \square \mathcal{R} \} g_{\mu \nu} 
+ \frac{1}{8 \pi} \mathcal{R}_{\mu \nu}   \label{eq2.7}
\end{eqnarray}

One can separate this tensor into anomalous $\langle T^{A}_{\mu \nu} \rangle$ and traceless $\langle T^{T}_{\mu \nu} \rangle$ part, $ \langle T_{\mu \nu} \rangle = \langle T^{A}_{\mu \nu} \rangle + \langle T^{T}_{\mu \nu} \rangle$. These two parts should be conserved separately $\nabla_{\mu}\langle T^{A\mu \nu} \rangle=\nabla_{\mu}\langle T^{T\mu \nu} \rangle=0$, as they have quite different origins. This is also in agreement with Bardeen \cite{Bardeen:2014uaa} and Christensen and Fulling \cite{ PhysRevD.15.2088} as they separated different parts of the energy momentum tensor including the conserved anomalous part.

In this note we only consider the effect of anomalous part. The traceless part  is responsible for the Hawking radiation effect \cite{Bardeen:2014uaa}. For the anomalous part we have, $\rho^{A}=-\langle T^{At}_{\ \ t} \rangle$, $p_{r}^{A}=\langle T^{Ar}_{\ \ r} \rangle$, and $p_{\perp}^{A}=\langle T^{A\theta}_{\ \ \theta} \rangle=\langle T^{A\phi}_{\ \ \phi} \rangle$.

Furthermore, Dymnikova and Bardeen also indicated that zero point energy \cite{Dymnikova:2001fb, Dymnikova:2003vt} including the conformal anomaly part \cite{Bardeen:2014uaa} (see also \cite{PhysRevD.56.2180, Ansoldi:2008jw}) should respect all of the symmetries of the Schwarzschild geometry which contains the invariance of Schwarzschild curvature tensor under radial boosts (an infalling observer in this geometry cannot measure the radial component of his velocity with respect to the zero point energy). So in this case, symmetry of the anomalous source is reduced from the full Lorentz group in flat space-time to the Lorentz boosts in the radial direction only. Therefore we obtain, $-\rho^{A}=p_{r}^{A}$, and from trace equation, $2p_{\perp}^{A}=\langle T_{\rho}^{\rho} \rangle - 2p_{r}^{A}$.
So we have,
\begin{eqnarray}
-\rho^{A}=p_{r}^{A}=\frac{1}{r^{4}}\int_{+\infty}^{r} r'^{3}\langle T_{\rho}^{\rho} \rangle dr' .\label{eq2.8}
\end{eqnarray}

In order to obtain the quantum corrected metric, one needs to solve Einstein equation with quantum sources ($\rho^{A}, p^{A}_{r}, p^{A}_{\perp}$). Exact solution of this equation is a challenging task since it is nonlinear and involves source terms with different higher powers of curvature (\ref{eq2.2}). In addition, as coefficients of higher order terms are a superposition of different curvature terms it is extremely difficult to guess the recurrence relation. Intuitively, these terms grow along with the curvature and create inner horizon at the point where they start to dominate the Einstein equation. In order to give an example for more clarification, let us consider the Reissner-Nordstr\"{o}m black hole where we have inner horizon. In this example inner horizon is created in a place where the fields which are on the other side of the Einstein equation ($G_{\mu\nu} = 8 \pi t_{\mu\nu} = 8 \pi (F_{\mu\rho} F_{\nu}^{\rho} - \frac{1}{4} g_{\mu\nu} F_{\rho \sigma} F^{\rho \sigma})$) dominate the Einstein tensor $G_{\mu\nu}$. Evidently, the exact solution approaches the Schwarzschild vacuum at infinity where the source terms vanish. So, the only natural approach is perturbation around the Schwarzschild vacuum. Accordingly in the Schwarzschild background the energy momentum  $ \langle T_{\mu \nu} \rangle $ takes the form,
\begin{eqnarray}
\langle T_{\mu}^{\nu} \rangle = -\frac{3c_{A}}{4 \pi}\left( \begin{smallmatrix} 1&0&0&0\\ 0&1&0&0\\ 0&0&-2&0\\ 0&0&0&-2 \end{smallmatrix} \right)\frac{l_{p}^{2} M^{2}}{r^{6}}
\end{eqnarray}

Inside the shell $V^{-}$ the metric is taken to be flat and for the region outside $V^{+}$, we solve the new Einstein equation with above  $ \langle T_{\mu \nu} \rangle $.

Let us assume the following form consistent with rotational symmetry, for the new metric in exterior region $V^{+}$,
\begin{eqnarray}
ds^{2}=- e^{\nu} dt^{2} + e^{\lambda} dr^{2} + r^{2}(d\theta^{2}+ \sin^{2}\theta d\phi^{2})
\end{eqnarray}

If we denote derivatives with respect to $t$ and $r$ by dot and prime, respectively, then the three independent components of the Einstein equations ($ G_{\theta}^{\theta}=G_{\phi}^{\phi}$ are dependent on the others) in first order of $\hbar$ are \cite{carroll2003spacetime, Antoci:2001hp},

\begin{eqnarray}
G_{t}^{t}=-e^{-\lambda}(\frac{\lambda^{\prime}}{r}-\frac{1}{r^{2}})-\frac{1}{r^{2}}=-\frac{6c_{A}l_{p}^{2} M^{2}}{r^{6}},
\end{eqnarray}
\begin{eqnarray}
G_{t}^{r}=e^{-\lambda}r^{-1} \dot{\lambda}=0.
\end{eqnarray}
\begin{eqnarray}
G_{r}^{r}=e^{-\lambda}(\frac{\nu^{\prime}}{r}+\frac{1}{r^{2}})-\frac{1}{r^{2}}=-\frac{6c_{A}l_{p}^{2} M^{2}}{r^{6}}.
\end{eqnarray}

Subtracting the first and second equation gives,
\begin{eqnarray}
\lambda^{\prime}+\nu^{\prime}=0,
\end{eqnarray}
and integration gives,
\begin{eqnarray}
\lambda+\nu=h(t),
\end{eqnarray}
where $h(t)$ is an arbitrary function of integration and  $\lambda$ is  a function of $r$. Hence we will have,
\begin{eqnarray}
(r e^{-\lambda})^{\prime}=1-\frac{6c_{A} l_{p}^{2} M^{2}}{r^{4}}. 
\end{eqnarray}
 where upon integration we obtain,
\begin{eqnarray}
e^{\lambda}=(1-\frac{2M}{r}+\frac{2c_{A} l_{p}^{2} M^{2}}{r^{4}})^{-1}
\end{eqnarray}

Finally, redefining the time coordinate we obtain the line element ($f(r)=e^{\nu}=e^{-\lambda}$) as,
\begin{eqnarray}
ds^{2}=-(1-\frac{2M}{r}+\frac{2c_{A} l_{p}^{2} M^{2}}{r^{4}}) dt^{2}
+ (1-\frac{2M}{r}+\frac{2c_{A} l_{p}^{2} M^{2}}{r^{4}}) ^{-1}dr^{2} \nonumber \\
+ r^{2}(d\theta^{2}+ \sin^{2}\theta d\phi^{2}) \label{eq2.17}
\end{eqnarray}

There are two horizons namely $r_{+}$ as outer horizon and $r_{-}$ as inner horizon given by,
\begin{eqnarray}
r_{+} \simeq \frac{2 M}{1+\left( \frac{R_{0}}{2M}\right)^{3} } \ \ \ \ \ \ \ \ \ \ \ \ \ \ \ \ \ \ \ \ \ \ \ \ \ \ \ \ \ \ \  \nonumber \\
r_{-} \simeq \frac{R_{0}}{1-\frac{R_{0}}{6 M}}, \ \ \ \ \ \ \ \ \ R_{0}=\left( c_{A} \frac{M}{M_{p}} \right)^{\frac{1}{3}} l_{p} \label{eq2.19}
\end{eqnarray}

\begin{figure}[t]
\centering
\includegraphics[width=0.75\textwidth]{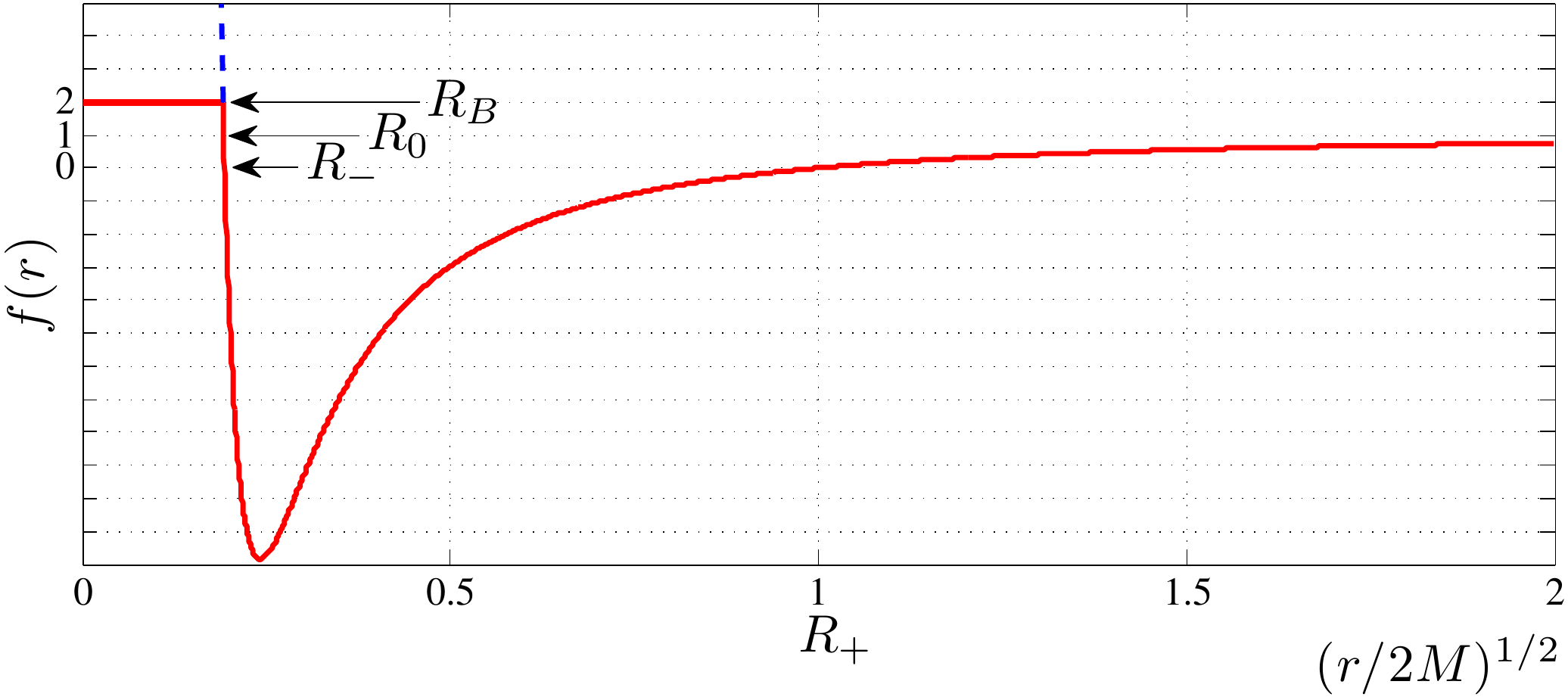}
\caption{\label{fig_1} Quantum corrected metric of Schwarzschild black hole ($f(r)=1-\frac{2M}{r}+\frac{2c_{A}l_{p}^{2} M^{2}}{r^{4}}$) at shell's minimum radius $R_{B}$ (inside the shell is flat region). Note that in the plot $R_{B}<R_{0}$ and as a result $m>R_{-}$. Dashed line represents the continuation of the quantum corrected Schwarzschild solution inside the shell. Numerical value is $c_{A}l_{p}^{2}/(2M)^{2}=2(R_{0}/2M)^{3}=10^{-4}$.}
\end{figure}

We see that the outer horizon $r_{+}$ is slightly smaller than Schwarzschild radius $2M$ and the inner horizon\footnote{As the equation (\ref{eq2.19}) shows inner horizon is created for positive $c_{A}$. For vector particles $c_{A}$ is negative, but higher order corrections stay positive inside the core such that the final result does not change.} $r_{-}$ is slightly bigger than $R_{0}$ where it is defined by $f(R_{0})=1$ (see figure \ref{fig_1}).

Note that for $M=\sqrt{\frac{32}{27}c_{A}} M_{p}$ the factor  $f(r)$ will have a double root and there will be an extremal black hole. For $M<\sqrt{\frac{32}{27}c_{A}}M_{p}$ the solution  will have a naked singularity and no horizon is formed. This value of $M$ sets a lower bound for the black hole mass. Interestingly its Hawking temperature will vanish.

The quantum corrected temperature of this black hole given by surface gravity  is,
\begin{eqnarray}
T_{H_{+}}=\frac{\kappa_{+}}{2 \pi}=\frac{1}{8 \pi M} \left( 1 - 2 \left( \frac{R_{0}}{2M}\right)^{3} \right).
\end{eqnarray}

Using the quantum modified metric we first obtain the geodesics of an infalling particle. We assumed that in this process it will be in region $V^{+}$ (outside the collapsing shell and not striking the shell) in all its trajectory.
In a gravitational field, equation of motion of this particle is given by, 
\begin{equation}
\frac{d^{2}x^{\mu}}{d\tau^{2}} + \Gamma^{\mu}_{\nu \rho} \frac{dx^{\nu}}{d\tau}\frac{dx^{\rho}}{d\tau} =0,
\end{equation}
where $m$ and $\omega$ are mass and energy of the particle respectively. Solution of this equation gives the geodesics of infalling massive particle \cite{9780511617676},
\begin{equation}
m^{2} \left(\frac{dR}{d\tau} \right)^{2} + V(R)=\omega^{2}
\end{equation}

where $V(R)=f(R)(m^{2}+J^{2}/R^{2})$ displays effective potential of black hole for classical particle. In this equation $R$ represents the position of the particle. Later we use it as radius of the collapsing shell. $\tau$  is time in inertial frame and $\omega$ is energy of this particle with angular momentum $J$.

Due to quantum corrections to the metric the potential barrier prevents the particle from falling into the singularity. This particle then moves in opposite direction as outgoing particle.
Setting $\dot R= \frac{dR}{d\tau} = 0$ gives the stopping point $f(R_{S})=(\omega / m)^{2}$ of this infalling radial particle ($J=0$). For $\frac{\omega}{m}<<(\frac{M}{M_{P}})^{1/3}$, this point is given by,
\begin{equation}
R_{S} \simeq \frac{R_{0}}{1+ \frac{R_{0}}{6M} \left( (\frac{\omega}{m})^{2} - 1 \right)}
\end{equation}
For a particle crossing the shell it can enter the flat region $V^{-}$ inside the shell.

Now let us determine the properties  of gravitational collapse or time evolution of R.  We assume also that the shell is made of pressureless matter. The dynamics of the radius is dictated by the matching condition of the inner and outer metrics. This matching condition has been known for a long time \cite{Israel:1966}.

Let us take a time-like hypersurface $\Sigma$, which divides the Riemannian space-time into two four-dimensional regions $V^{-}$ and $V^{+}$. On $\Sigma$, the intrinsic co-ordinates $\xi^{a}$, and in $V^{-}$ and $V^{+}$, the mutually independent coordinate charts $x_{-}^{a}$ and $x_{+}^{a}$ are introduced. 

In the exterior region $V^{+}$ the metric is given by  (\ref{eq2.17}) and the region $V^{-}$, interior to the shell is flat,
\begin{eqnarray}
(ds^{2})^{-}=- dt_{-}^{2} + dr^{2}+ r^{2}(d\theta^{2}+ \sin^{2}\theta d\phi^{2}) \label{eq2.20}
\end{eqnarray}

The interior and exterior co-ordinates do not necessarily join continuously ($t_{+} \neq t_{-}$ on $\Sigma$). The induced metric on the surface of the shell should be the same computed from either side, of course up to a diffeomorphism. In addition, if the shell is freely falling, its motion is determined by matching the extrinsic curvature of the geometry across the shell.

Now that we have found the quantum correction to the Schwarzschild solution, we can proceed to study the problem of collapse. The motion of the shell is expressed by the equation $r=R(\tau)$ where $\tau$ is the co-moving time. The line element of the shell, induced both by the exterior element (\ref{eq2.17}) and the interior element (\ref{eq2.20}) is given by
\begin{eqnarray}
ds^{2}=- d\tau^{2} + R(\tau)^{2}(d\theta^{2}+ \sin^{2}\theta d\phi^{2})
\end{eqnarray}
The derivative with respect to $\tau$ is denoted by a dot.

The energy momentum tensor for a spherically-symmetric shell of pressureless dust following a radial trajectory $R(\tau)$ and with density $\sigma$ is given by
\begin{eqnarray}
t^{\mu \nu} = \sigma u^{\mu}u^{\nu} \delta(r-R(\tau))
\end{eqnarray}
where $u_{\mu}$ is the velocity of the shell, satisfying $u_{\mu}u^{\mu}=-1$. It fails to obey  local conservation law because of the conformal anomaly, vacuum polarization, and Hawking radiation effects.

The mechanical properties of matter are described by the surface energy-momentum tensor $t_{\mu \nu}$. An observer which moves with an element of the shell finds that the momentum of matter is always restricted to the surface of the shell, so that the normal components of momentum and momentum current remain zero. This means that
\begin{eqnarray}
t_{\mu \nu}n^{\nu}=0, \ \ \ \ \ \ on \ \ \Sigma, \nonumber \\
t_{\mu \nu}=0 \ \ \ \ \ outside \ \ \Sigma.
\end{eqnarray}
where,
\begin{eqnarray}
u^{\mu}_{\pm}=(\dot{t}_{\pm}, \dot{R}, 0, 0), \ \ \ \  n_{\mu}^{\pm}=(-\dot{R}, \dot{t}_{\pm}, 0, 0)
\end{eqnarray}

The surface energy-momentum tensor $t_{\mu \nu}$ is therefore represented by the intrinsic tensor $S_{ab}$ for an observer on $\Sigma$. For ideal fluid, the intrinsic energy-momentum tensor for a spherically-symmetric shell of pressureless dust has the form
\begin{eqnarray}
S_{ab}=\sigma u_{a} u_{b}
\end{eqnarray}

Here the  Latin indices run through 0, 1, 2. The metric of the time-like hypersurfaces has the signature $(-,+,+)$.

Thereby, $\sigma$  is the rest mass density of the surface  of the shell. 
The energy-momentum tensor in the region $V^{+}$ outside the shell originated exclusively from the vacuum energy induced by quantum fluctuations $\langle T_{\mu \nu} \rangle$ coupled to strong gravitational field similar to  the  effects known as vacuum polarization, conformal anomaly, and particle creation in Hawking radiation.

The energy-momentum tensor inside the shell (flat region) originates exclusively from the vacuum energy called Casimir energy. Since inside the shell the geometry is flat and anomaly vanishes its contribution is neglected.

Now we proceed to apply the matching conditions; first we consider the continuity of the metric $[g_{ij}] = 0$ which  implies  $f_{\pm}^{2}(dt_{\pm}/d\tau)^{2}=(dR/d\tau)^{2}+f_{\pm}$ and $R^2 = R(\tau)^{2} 
$ and secondly the discontinuity of the extrinsic curvatures  $[K_{ij}] \neq 0$ due to  nonvanishing
surface stress-energy. The extrinsic curvatures on $V^{\pm}$ \cite{poisson} are
\begin{eqnarray}
K_{\pm}^{\tau}{}_{\tau}=\dot{\beta_{\pm}}/\dot{R},
\end{eqnarray}
\begin{eqnarray}
K_{\pm}^{\theta}{}_{\theta}=K_{\pm}^{\phi}{}_{\phi}=\beta_{\pm}/R,
\end{eqnarray}
where
\begin{eqnarray}
\beta_{+}=\sqrt{\dot{R}^{2}+1-\frac{2M}{R}+ \frac{2c_{A} l_{p}^{2} M^{2}}{R^{4}} }
\end{eqnarray}
\begin{eqnarray}
\beta_{-}=\sqrt{\dot{R}^{2}+1}
\end{eqnarray}

The surface stress-energy is defined by
\begin{eqnarray}
S_{ab}=-([K_{ab}]-[K]h_{ab})
\end{eqnarray}
where $h_{ab} = g_{ab} + u_{a}u_{b}$ is an induced metric on $\Sigma$.

The surface stress-energy can be  evaluated,
\begin{eqnarray}
-\sigma = S^{\tau}_{\tau}=\frac{1}{4\pi R}(\beta_{+}-\beta_{-}), \label{eq2.31}
\end{eqnarray}
\begin{eqnarray}
0=S^{\theta}_{\theta}=\frac{1}{8\pi R}(\beta_{+}-\beta_{-})+\frac{1}{8\pi \dot{R}}(\dot{\beta}_{+}-\dot{\beta}_{-}). \label{eq2.32}
\end{eqnarray}

The second equation can be integrated immediately, giving $(\beta_{+}-\beta_{-})R=constant$. Substituting this into the first equation yields
\begin{eqnarray}
4 \pi R^{2} \sigma \equiv m  \label{eq2.33}
\end{eqnarray}

where m is the constant of integration and $(\beta_{-}-\beta_{+})=m/R$.

The first equation states that $m$, the shell's rest mass, stays constant during the evolution and the second one gives the dynamics of the radius of the shell,
\begin{eqnarray}
M=m\sqrt{\dot{R}^{2}+1}-\frac{m^{2}}{2R}+  \frac{c_{A} l_{p}^{2} M^{2}}{R^{3}} \label{eq2.34}
\end{eqnarray}
This equation has a nice physical interpretation. The first term on the right-hand side is the shell's relativistic kinetic energy, including rest mass. The second term is the shell's binding energy, the work required to bring all its parts together. The last term is the energy eaten up by the vacuum. They are extracted from the initial energy of the matter $M$ at long past where it is dilute and the background is flat.
The sum of these is the total (conserved) energy, and this is equal to the shell's gravitational mass M at infinity. Equation (\ref{eq2.34})  illustrates the general statement that all forms of energy contribute to the total gravitational mass.

Equations (\ref{eq2.33}) and (\ref{eq2.34})  are the shell's equations of motion. (\ref{eq2.34}) can be alternatively written in the form,
\begin{eqnarray}
\dot{R}^{2}+V_{eff}(R)=0,\label{eq2.36}
\end{eqnarray}
where the effective potential is given by
\begin{eqnarray}
V_{eff}(R)=1-\frac{1}{m^{2}} \left( M+\frac{m^{2}}{2R} -  \frac{c_{A} l_{p}^{2} M^{2}}{R^{3}} \right)^{2}.\label{eq2.37}  \nonumber \\
\end{eqnarray}

We divide the problem into two cases ($m \leq R_{-}$ and $m > R_{-}$).
First let us investigate the case $m \leq R_{-}$. Considering the equation (\ref{eq2.36}) and (\ref{eq2.37}) and setting $\dot{R}^{2}=-V_{eff}(R)=0$ gives the bounce radius $R_{B} $ which is between $R_{0} $ and  $  R_{-}$. Therefore we have,
\begin{eqnarray}
(1-\frac{m}{R_{B}})^{2}=f(R_{B}), \ \ \ \  m \leq R_{-}, \ \ \ \  R_{0} \leq R_{B} \leq R_{-}.
\end{eqnarray}

\begin{figure}[t]
\centering
\includegraphics[width=0.75\textwidth]{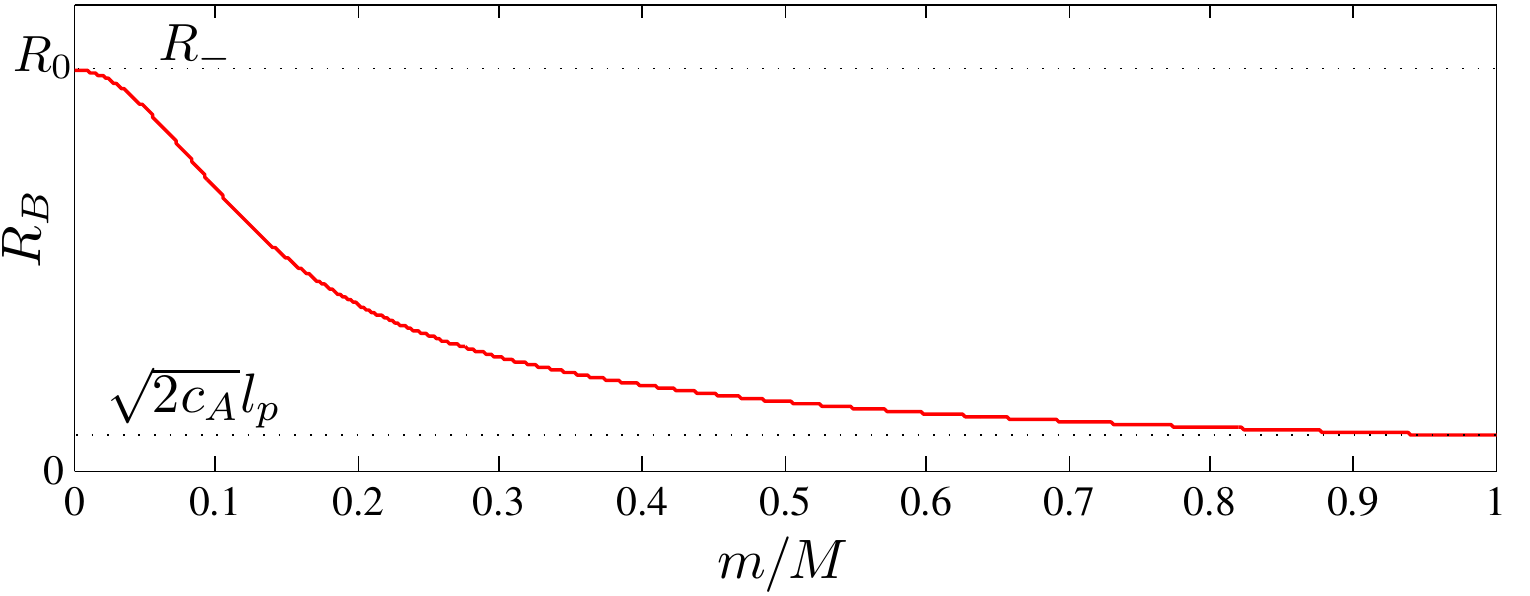}
\caption{\label{fig_2}Plot of bounce radius as a function of rest mass $m$. Numerical value is $c_{A} l_{p}^{2}/(2M)^{2}=2(R_{0}/2M)^{3}=1.25 \times 10^{-8}$.}
\end{figure}

In this case for infinitely light shell $m\rightarrow 0$, bounce radius tends to $R_{B} \rightarrow R_{0}=\left( c_{A} \frac{M}{M_{p}} \right)^{\frac{1}{3}} l_{p}$ (this radius is inside the inner horizon of the black hole), and for $m=R_{-}$ bounce radius becomes $R_{B}=R_{-}$ (see figure \ref{fig_2}).

For the second case ($m>R_{-}$), effective potential for this equation is the same as the former, (\ref{eq2.37}). Therefore bounce radius (see figure \ref{fig_2}) in this case which is between $\sqrt{2c_{A}} l_{p}<R_{B}<R_{-}$ is given by,
\begin{eqnarray}
(1-\frac{m}{R_{B}})^{2}=f(R_{B}), \ \ \ \ \ \ \  m > R_{-}.
\end{eqnarray}

\begin{figure}[t]
\centering
\includegraphics[width=0.4\textwidth]{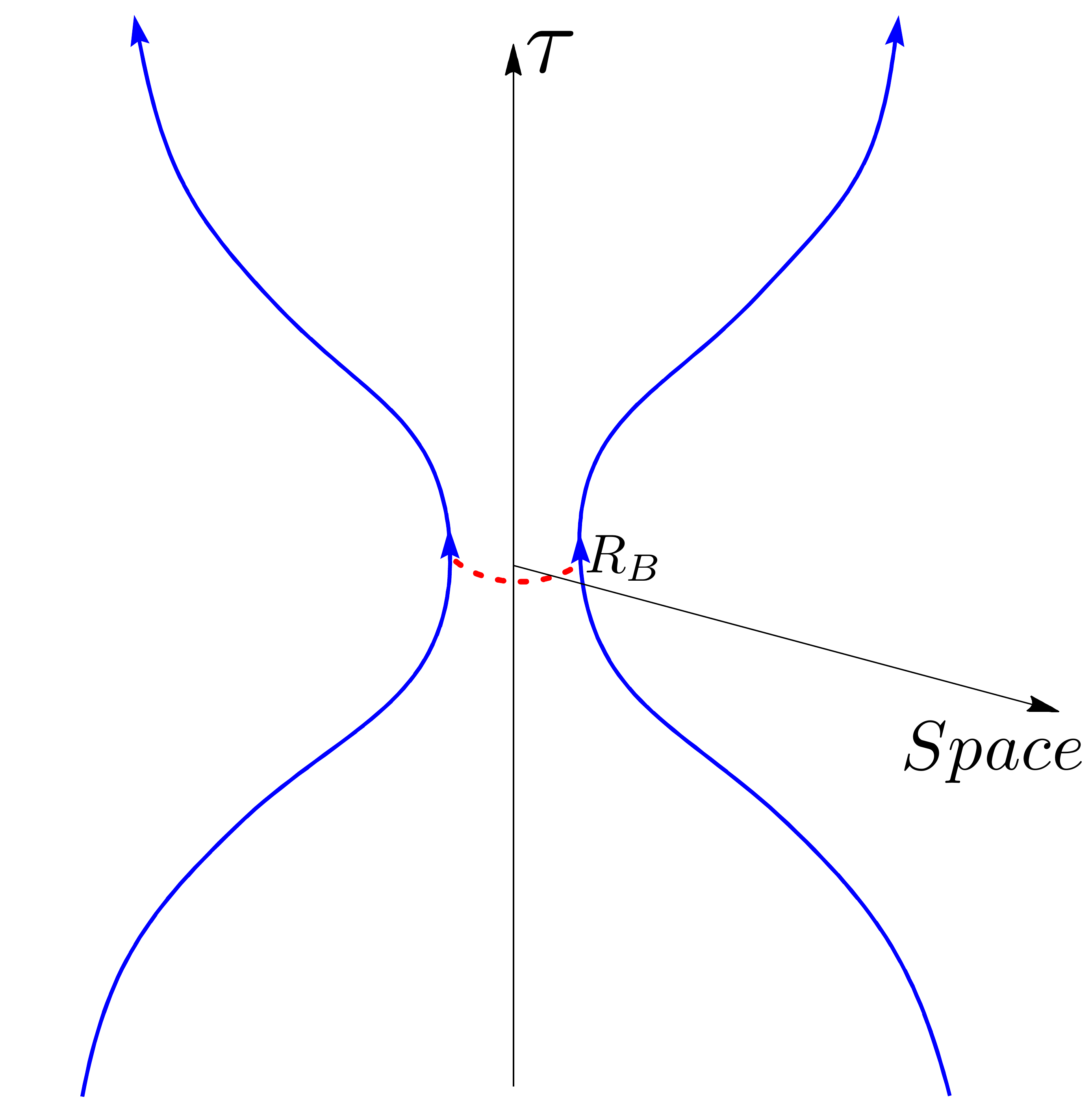}
\caption{\label{fig_3}Spherical gravitational collapse and its expansion in co-moving coordinates.  Dashed line indicates bounce radius $R_{B}$ which is inside the inner horizon $R_{-}$.}
\end{figure}

One can determine the trajectory of this shell in terms of its radius and time in an arbitrary frame which we choose to be the co-moving one. For a co-moving frame (see figure \ref{fig_3}) this trajectory is given by,
\begin{eqnarray}
\tau= 
\pm \int \frac{2m}{R} \frac{dR}{\sqrt{((1-\frac{m}{R})^{2} - f(R) )((1+\frac{m}{R})^{2} -f(R) )}}
\end{eqnarray}
where$ +$  and $-$ refer to direction of the shell's trajectory.

For a co-moving frame shown in the figure \ref{fig_3} we see that the shell expands after reaching the bounce radius.
It is worth considering the trajectory in terms of a non-singular metric such as ingoing Eddington-Finkelstein coordinates,
\begin{eqnarray}
ds^{2}=-f(r) dv^{2} + 2 dv dr 
+ r^{2}(d\theta^{2}+ \sin^{2}\theta d\phi^{2})
\end{eqnarray}
so the ingoing time $\nu=t+r_{\ast}$ is given by,
\begin{eqnarray}
v= 
 \int \frac{dR}{f(R)} \left( 1 - \frac{1 - f(R) - \frac{m^{2}}{R^{2}}}{\sqrt{( (1-\frac{m}{R})^{2} - f(R) )( (1+\frac{m}{R})^{2} - f(R) )}}  \right)
\end{eqnarray}
It can be easily verified that the above ingoing trajectory is regular at $r_{\pm}$.
After the bounce which is inside the inner horizon, the shell expands and the trajectory becomes outgoing ($\dot{R}>0$). Therefore we have,
\begin{eqnarray}
v=
\int \frac{dR}{f(R)} \left( 1 + \frac{1 - f(R) - \frac{m^{2}}{R^{2}}}{\sqrt{( (1-\frac{m}{R})^{2} - f(R) )( (1+\frac{m}{R})^{2} - f(R) )}}  \right)
\end{eqnarray}
It can be seen that the outgoing trajectory is singular at $r_{\pm}$. Therefore the outgoing shell cannot exit the horizon as long as the trapping surfaces exist i.e. the coordinate singularity remains in the metric. In other words the shell cannot exit the horizon within a finite time.

\begin{figure}[t]
\centering
\includegraphics[width=0.30\textwidth]{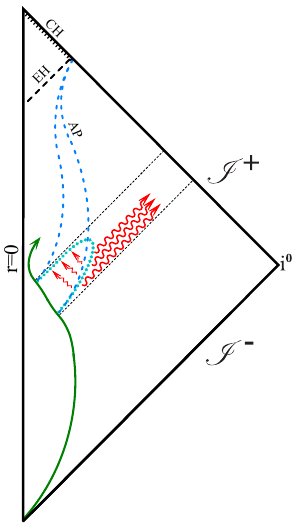}
\caption{\label{fig_4}This figure represents the penrose diagram of gravitational collapse of a shell and its expansion. The collapsing shell is indicated by solid line dividing the interior flat region and exterior quantum modified Schwarzschild geometry. In our scenario trapping horizons (the external evaporating one, and the internal shrinking one) uniting and forming a stable extremal black hole at future infinity are represented by dashed line. Thick straight dashed line represents event horizon (EH) \cite{Bonanno:2006eu,Myung:2006mz}. Apparent horizons (AH) are indicated by either dashed and dotted curves \cite{Bonanno:2006eu,Myung:2006mz,Hossenfelder:2009fc,Alesci:2011wn}. On the other hand, the dotted straight line represents the Cauchy horizon (CH) \cite{Chen:2014jwq} and our ignorance on the resolution of the instability and blueshift effect. The two trapping horizons (the external evaporating one, and the internal accreting one) according to the Hayward scenario represented by dotted line \cite{Hayward:2005gi, Rovelli:2014cta,DeLorenzo:2014pta}. Hawking radiation (outgoing and ingoing) is represented by ingoing and outgoing arrows. Straight thin dashed lines indicate the two (inner and outer) horizons for non-evaporating black hole.}
\end{figure}

Hawking radiation will result in decrease of the mass and hence slow shrinking of the outer and inner horizons (see figure \ref{fig_4}). Similar to Vaidya solution this can be  represented in the metric by time dependence of the gravitational mass $M(t)$. Finally the two horizons merge at $R_{+}=R_{-}=R_{H}$ creating an extremal black hole. In this case the temperature of the black hole vanishes and the black hole as it has no conserved charge becomes stable. According to the third law of black hole thermodynamics this is not achieved in finite time (steps).

\section{\label{Conclusion}Conclusion and discussion}

We have shown that QFT corrections due to trace anomaly, as terms  induced in the Einstein equation can provide mechanism in the Schwarzschild black hole  that are strong enough  to obstruct the formation of singularity of a collapsing shell far from the center at the radius of order $ (M/M_{p})^{1/3} l_{p} $. This radius can be much larger than the Planck length. We used the benefit of the exactness of trace anomaly that allows us to track the regions beyond the reliability of semi-classical approximations. Without invoking Quantum Gravity in any of its forms and only using QFT effects we have addressed the fate of the collapsing shell. The effect is similar to the Casimir energy effects where the difference in the vacuum energy density results in a pressure on the boundary surface of a closed region. Also as can be seen from the (\ref{eq2.36}) in the ingoing phase it is clearly similar to the classical motion of a particle in the potential $V_{eff}(R)$.

The backreaction of the induced QFT effects has strong consequences for the solution; it changes the Schwarzschild radius and creates a new inner horizon, it also prevents the formation of a black hole with mass less than $\sqrt{\frac{32}{27}c_{A}}M_{p}$. Solutions with mass equal $\sqrt{\frac{32}{27}c_{A}}M_{p}$ form extremal Schwarzschild black hole with zero temperature. The extremal black holes  are  free of any charges hence will have only gravitational interaction. We expect it be the final state of a black hole after Hawking radiation which will be reached asymptotically and becoming stable due the third law of black hole thermodynamics (see figure \ref{fig_4}). This changes have profound effects on the process of collapse and prevents the formation of singular object. The turning point, the bounce radius, is inside the inner horizon. Moreover the quantum corrections to the background energy modifies the surface gravity. The surface area of the black hole is reduced and in turn this will change the Hawking temperature.

In this background the shell reaches to the point inside the inner horizon where the repulsive force due to vacuum energy overcomes the gravitational attraction and the process of the collapse reverses.

The turning point has the radius, $R_{B}$ in accordance  with the  conservation of energy; all the kinetic energy is transferred to the empty region out of the shell, mainly between the shell and the Schwarzschild radius. Energy is extracted from the shell in two forms, the Hawking radiation which is emitted and goes out to infinity, and the energy stored in the background as the zero point energy and grows with the second power of the curvature. We have explored the immediate consequence of the second type that arises from the trace anomaly.

After reaching the bounce radius the matter expands back. From the point of view of the external observer, due to large time dilation this process takes a very long time in comparison to the evaporation time. Hence for a long period the star resembles the standard black hole.

Among the sources that can strongly affect the result is the traceless part, which includes Hawking radiation, not studied in detail in this paper. It will affect the lifetime of the black hole and consequently the expansion time. In general it also contributes to damping of the collapse and giving rise to a loss of original mass, it will increase the bounce radius.

As discussed earlier the energy density of the fields induced from trace anomaly play the essential role in the obstruction of the singularity. In the lowest order of perturbation they are represented by $1/r^4$ term in the metric components. If it's coefficient is positive, as in most cases, it results in the obstruction of the formation of the singularity. They have given positive coefficient as we go further. The next term we encounter is of the order of $1/r^7$ and then $1/r^{10}$ and etc. In this case the formation of the inner horizon happens before the higher powers of $1/r$ are turned on. But there may be case where the coefficient of this term is negative. In such cases we have to go further in our perturbation series and consider higher powers of $1/r$. Taking into account of quadratic terms in trace anomaly, we have considered the series up to $1/r^{22}$. The formation of the inner horizon happens at the first positive term. Although it is a challenging task for obtaining the exact solution and giving general statement, from our primary investigations and also numerical calculations the contribution to the effective term is positive inside the core. Classification of different field combinations leading to different first positive terms is under investigation and will presented elsewhere.

There are also other scenarios (see figure \ref{fig_4}); According to the scenario proposed by Hayward \cite{Hayward:2005gi} (see \cite{Rovelli:2014cta, Bardeen:2014uaa, Barrau:2014hda,DeLorenzo:2014pta}), the endpoint of evaporation defined by the disappearance of trapped surfaces, occurs when the outer and inner section of the trapping horizon unite and disappear. In this time measured by outside observer this shell emerges. For a non-evaporating (non-physical) scenario the shell emerges at past infinity \cite{PhysRev.153.1388} in another universe \cite{Bambi:2013gva}.

We have also shown that a similar phenomenon happens for an infalling particle in this quantum corrected background. The geometry acts like an effective potential preventing the particle reaching the center.

The phenomena can be extended to other black holes such, Reisner-Nordstr\"{o}m and Kerr without much toil. Application of this method to other black holes such as dilaton, AdS, Reissner-Nordstr\"{o}m, and Kerr is under investigation.

Similar analysis can modify the behavior of any singularity of curvature including big bang solutions. QG considerations predict resolution of big bang singularity at  a bounce radius, called big bounce \cite{Ashtekar:2006rx, Ashtekar:2005qt, Modesto:2005zm}. We expect similar effect from QFT corrections.

The consequence of our result for the problem of information loss is also consistent with the observation made for Planck star \cite{Rovelli:2014cta,DeLorenzo:2014pta}. So quantum corrections may also provide a way to solve the information loss problem.
The above issues are under investigation and will be addressed in detail elsewhere.

\acknowledgments

The authors are grateful to R. Mansouri and J. T. Firouzjaee for useful conversations. The authors would like to thank an anonymous referee for the valuable comments and suggestions to improve the clarity and the quality of this paper.


\bibliography{Paper-Abedi}

\end{document}